\documentclass[11pt,a4paper]{article}
\pdfoutput=1
\usepackage{empheq}
\usepackage{cases}
\usepackage{a4wide}
\usepackage{amsfonts}
\usepackage{amssymb}
\usepackage{amsmath,bm}
\usepackage{amsmath}
\usepackage{graphicx}
\usepackage{mathtools}
\usepackage{booktabs}
\usepackage{array}
\usepackage{color}
\usepackage{ulem}
\usepackage[small,bf]{caption}
\usepackage{cite}
\usepackage[usenames,dvipsnames]{xcolor}
\usepackage{subfigure}
\usepackage{verbatim}

\allowdisplaybreaks

\numberwithin{equation}{section}

\newcounter{mysubequation}[equation]

\DeclarePairedDelimiter\bra{\langle}{\rvert}
\DeclarePairedDelimiter\ket{\lvert}{\rangle}
\DeclarePairedDelimiterX\braket[2]{\langle}{\rangle}{#1 \delimsize\vert #2}

\begin{document}
\begin{titlepage}

\begin{center}
{
\bf\LARGE Detection Prospects of Millicharged Dark Matter\\[0.3em]
in Unconventional Interferometer
}
\\[8mm]
Chrisna~Setyo~Nugroho\footnote[1]{setyo13nugros@ntnu.edu.tw}  
\\[1mm]
\end{center}
\vspace*{0.50cm}

\centerline{\it Department of Physics, National Taiwan Normal University, Taipei 116, Taiwan}
\vspace*{1.20cm}

\begin{abstract}
\noindent
We propose a novel idea to discover millicharged particles
(mCPs) captured by the earth during its existence. It has been
demonstrated that the mCPs accumulation inside the earth leads to 
the enhancement of its number density much larger than the
corresponding virial density.   
We propose to utilize an unconventional laser interferometer
to probe these earth-bound mCPs through the detection of the photons's phase shift. We demonstrate that, for mCPs mass in the
range between $1$ GeV to $10^{12}$ GeV, the sensitivity of
probing their fractional electric charge $\epsilon$ could
reach as low as $10^{-12}$ to $10^{-6}$ provided that mCPs number density is greater than $1~\rm{cm^{-3}}$ and the interferometer operates at the Heisenberg limit.
\end{abstract}

\end{titlepage}
\setcounter{footnote}{0}

\section{Introduction}

The existence of non-luminous matter dubbed as dark matter (DM) has been haunted multidisciplinary fields such astrophysics, cosmology, as well as particle physics for decades. 
Based on existing observations, ranging from galactic rotational
curve to gravitational lensing, one concludes that it gravitationally interacts
with ordinary matter while its other detail properties remain unknown. One of the popular and well motivated DM candidates is millicharged particles proposed to explain the absence of magnetic monopole. Although mCPs have non-zero electric charge, they are viable dark matter candidates with suitable observed abundant as has been shown in~\cite{Dvorkin:2019zdi,Creque-Sarbinowski:2019mcm,Nguyen:2022zwb,Plestid:2020kdm,Hu:2016xas}.

Many experimental attempts have been carried out to probe the existence of mCPs 
~\cite{Prinz:1998ua,Magill:2018tbb,Marocco:2020dqu,Ball:2020dnx,ArgoNeuT:2019ckq,Foot:2014uba,Foot:2014osa,Foot:2016wvj}. From astrophysical frontiers, the null results from
anomalous emission in stellar environments place strong constraints on
mCPs mass below MeV range~\cite{Davidson:2000hf,Chang:2018rso,Fiorillo:2024upk}. Recently, there are several proposed strategies to probe mCPs~\cite{Knapen:2017ekk,Blanco:2019lrf,Essig:2019kfe,Berlin:2019uco,Kurinsky:2019pgb,SENSEI:2020dpa,Griffin:2020lgd,Chen:2022abz}. Moreover, mCPs also escape the
dedicated dark matter direct search experiment such as XENON1T~\cite{XENON:2020rca}. 

Recent studies show that mCPs may lost its virial kinetic
energy and trapped inside the earth~\cite{Neufeld:2018slx,Pospelov:2020ktu}. For mCPs mass larger than 1 GeV, the earth's gravitational pull sinks mCPs to the earths's core
leading to the significant mCPs accumulation during the
earth's existence. 
Since mCPs have non-vanishing interaction with the photon, 
laser interferometer offers a suitable venue for terrestrial mCPs
search. When mCPs interact with the laser in one arm of the
interferometer, it would induce a phase shift on the laser to be detected at the
output port. However, the existing DM search proposals utilizing the laser interferometer employed at Gravitational Wave (GW) experiments~\cite{Tsuchida:2019hhc,Chen:2021apc,Lee:2020dcd,Ismail:2022ukp,Lee:2022tsw,Seto:2004zu,Adams:2004pk,PhysRevLett.114.161301,Arvanitaki:2015iga,Stadnik:2015xbn,Branca:2016rez,Riedel:2016acj,Hall:2016usm,Jung:2017flg,Pierce:2018xmy,Morisaki:2018htj,Grote:2019uvn} are not
suitable for mCPs search since both
of the interferometer arms are positioned at the same depth
underground leading to zero phase shift in the photon path.

To overcome this problem, we propose a phase measurement scheme based on the optical laser
experiment using unconventional interferometer which has a single
arm and single photon mode. This kind of interferometer is known in quantum optics literature, see, for instance~\cite{Ou:2017}. To probe earth-bound mCPs, the arm needs to be placed vertically from the earth's surface with 1 km arm length. We employ photon vacuum state as our input and output as well as the squeezed vacuum state to probe the phase
shift. We show that our proposal is sensitive to detect the mCPs
charge as low as $10^{-12}\,e$ for mCPs number density equals to 1 $\text{cm}^{-3}$, with $e$ denotes the magnitude of
electron charge. This is more  sensitive
than recent constraints on heavy mCPs mass regime given by
collider experiments as well as the projected limits set by the ion trap proposal~\cite{Budker:2021quh}.

The  paper is structured as follows: In section~\ref{sec:ebdm}, we
give a brief discussion on earth-bound mCPs relic. The interaction between mCPs and photons is studied in Section~\ref{sec:interaction}. We introduce an optical phase measurement based on single arm unconventional
interferometer in Section~\ref{sec:phase} and further present its
projected sensitivity to probe mCPs 
in section~\ref{sec:result}.
Our summary and conclusion are presented in Section~\ref{sec:Summary}.

\section{A Lightning Review of Earth-bound mCPs}
\label{sec:ebdm}

The scattering between dark matter and ordinary matter such as
atmosphere, earth's crust, as well as shielding materials has
been discussed in detail in~\cite{Emken:2019tni}. For mCPs, the corresponding scattering leads to significant accumulation of mCPs 
on earth as has been demonstrated in~\cite{Pospelov:2020ktu}. Here, we
summarize the important findings of~\cite{Pospelov:2020ktu} and employ the relevant results to our study.  
 
It has been shown that the evaporation of mCPs with $m_Q\gtrsim 1$ GeV from the earth can be neglected. As a result, mCPs are trapped in the earth with the corresponding average number density on earth given by 
\begin{align}
\label{eq:avedensity}
<n_Q> \simeq <n_Q^{cap}> \approx f_Q\left(\frac{t_\oplus}{10^{10}~\rm year}\right)\left(\frac{\rm GeV}{m_Q}\right)\left(\frac{3\times 10^{15}}{\rm cm^3}\right),
\end{align}  
where $f_Q$ and $t_\oplus$ are the fraction of mCPs with respect to the total local DM density and the age of the earth,
respectively. Moreover, thanks to the earth's gravitational pull, mCPs would sink underground reaching a terminal
velocity given by 
\begin{align}
\label{eq:vterm}
v_{\rm term} &= \frac{3m_Q g T}{m^2_{\rm rock}n_{\rm rock}<\sigma_T v^3_{\rm th}>}~~{\rm for}~~m_Q>m_{\rm rock}\,, \\
{} &= \frac{m_Q g}{3 n_{\rm rock}T}\left<\frac{v_{\rm th}}{\sigma_T}\right>~~{\rm for}~~m_Q<m_{\rm rock}\,.
\end{align}  
Here, $m_{\rm rock}$ and $n_{\rm rock}$ are the mass and
number density of terrestrial medium atom, respectively, while $g$ and $\sigma_{T}$ stand for the earth's gravitational
acceleration and transfer cross section between mCPs and 
terrestrial medium. Note that
the terminal velocity also depends on the thermal velocity of
mCPs after their thermalization in the atmosphere $v_{\rm th}$. Since the terminal velocity is smaller than the average virial mCPs velocity $v_{\rm vir}$, the so-called traffic jam effect enhances mCPs number density $n_{\rm tj}$ 
\begin{equation}
\label{eq:ntj0}
n_{\rm tj} = \frac{v_{\rm vir}}{v_{\rm term}}n_{vir}\,,
\end{equation}
where $n_{\rm vir}$ denotes the number density of galactic mCPs. Finally, the number density of mCPs underground $n_{\rm loc}$ has been calculated as 
\begin{equation}
\label{eq:ntj}
n_{\rm loc} = {\rm Max} \left(n_{\text{Jeans}}, {\rm Min}(n_{\rm tj}, \left<n_Q\right>)\right)\,.
\end{equation}
Here, $n_{\text{Jeans}}$ is the number density governed by the Jean's equation for a static, steady-state distribution of mCPs~\cite{Neufeld:2018slx}. It is worth to mention that such top-down accumulation after thermalization holds for mCPs with large enough $\epsilon$ \cite{Pospelov:2020ktu}.

On the other hand, mCPs with smaller $\epsilon$ could also accumulate underground before experiencing thermalization (bottom-up accumulation). In this case, the corresponding number density is given by \cite{Pospelov:2020ktu}

\begin{align}
\label{eq:nsmalleps}
n_{\text{loc}}(h) = \int^{\beta \gamma_{\text{max}}} d(\beta \gamma)\, \frac{d\Phi}{d(\beta \gamma)}\, \frac{\pi\, R^{2}_{\oplus} \, t_{\text{diff}}(d_{\text{pend}})}{\frac{4\pi}{3} (R^{3}_{\oplus}-(R^{3}_{\oplus}-d_{\text{pend}})^{3})} \, \frac{h}{d_{\text{pend}}}\,.
\end{align}
Here, $\frac{d\Phi}{d(\beta \gamma)}$ and $R_{\oplus}$ stand
for mCPs flux per boost factor interval $\beta \gamma$ and the
radius of the earth, respectively. Such mCPs could penetrate
the rock with penetration distance $d_{\text{pend}}(m_{Q},\epsilon,\beta \gamma)$. Furthermore, the diffusion time $t_{\text{diff}}$ required by such mCPs to reach the surface from the bottom depends on $d_{\text{pend}}$. The upper integration limit $\beta \gamma_{\text{max}}$ is obtained by equating $d_{\text{pend}} = R_{\oplus}$.

Following~\cite{Budker:2021quh}, we assume that all mCPs considered
here are
free and there is asymmetric mCPs with opposite charges
assigned to different species (like SM proton and electron)
such that there is no annihilation to prevent terrestrial
accumulation.
Furthermore, we focus on the case of heavy mCPs ($m_Q\gtrsim 1$ GeV) and take three
benchmark values of underground number density $n_{\rm loc}=$ $1~{\rm cm^{-3}}$, $10^3~{\rm cm^{-3}}$, and $10^6~{\rm cm^{-3}}$ in our study to compare with ion trap proposal, see Fig.2 of~\cite{Budker:2021quh}.
 
\section{mCPs and Photon Interaction}
\label{sec:interaction}

As mCPs lost their virial kinetic energy, they are trapped
inside the earth. To detect their existence with laser
interferometer, one needs to study the interaction between
non-relativistic charged particles with photons. The relevant Hamiltonian is given by
\begin{align}
\label{eq:hamiltontot}
H = H_{P} + H_{R} + H_{I} \,,
\end{align}
where $H_{P}$, $H_{R}$, and $H_{I}$ are the Hamiltonian for free charged
particles, free photon field, and the interaction between charged particles and the photons,
respectively. Their explicit expression are given by~\cite{cohen:1987}
\begin{align}
\label{eq:hamilall}
H_{P} &= \sum_{s} \frac{\vec{p}^{2}_{s}}{2\, m_{s}} + V_{\text{Coulomb}}\,,\\
H_{R} &= \sum_{i} \hbar \omega_{i} \left( \hat{a}^{\dagger}_{i} \hat{a}_{i} + \frac{1}{2}\right)\,,\\
H_{I} &= H_{I1} + H_{I2}\,,\\
H_{I1} &= - \sum_{s} \frac{\text{q}_{s}}{m_{s}} \, \vec{p}_{s} \cdot \vec{A}(\vec{r}_{s})\,,\\
H_{I2} &= \sum_{s} \frac{\text{q}^{2}_{s}}{2\,m_{s}} \left[ \vec{A}(\vec{r}_{s})\right]^{2},
\end{align}  
where $\vec{p}_{s}$, $m_{s}$, and $\text{q}_{s}$
stand for the momentum, the mass, and the electric charge of the  $s$-th charged particle, respectively. The operator $\hat{a}_{i}$ ($\hat{a}^{\dagger}_{i}$) denotes the annihilation (creation)
operator of the photon field for the i-th mode which obeys the
commutation relation $[\hat{a}_{i},\hat{a}^{\dagger}_{j} ] = \delta_{ij}$. In terms of these operators, the photon field can be written as~\cite{cohen:1987}
\begin{align}
\label{eq:photon}
\vec{A}(\vec{r}) = \sum_{i} \left[ \frac{\hbar}{2 \,\epsilon_{0}\, \omega_{i} L^{3}} \right]^{1/2} \left[\hat{a}_{i}\, \vec{\varepsilon}_{i} \, e^{\text{i} \vec{k}_{i} \cdot \vec{r}} + \hat{a}^{\dagger}_{i}\, \vec{\varepsilon}_{i} \, e^{-\text{i} \vec{k}_{i} \cdot \vec{r}} \right] \,.
\end{align}
We quantize the photon field $\vec{A}(\vec{r})$ in a box  of volume $L^{3}$ with a normalization condition $\vec{k} \cdot \vec{L} = 2\pi \,n$ with $n$ takes integer values.
Note that the wave number and the angular frequency of the photon are related via $ \omega = |\vec{k}|\, c$. 

When the photons interact with mCPs, they
experience the phase shift $\delta$ which is encoded in the
intercation Hamiltonian $H_{I} = H_{I1} + H_{I2}$. The first term $H_{I1}$ is
suppressed by mCPs velocity. It is also proportional
to $(\hat{a} + \hat{a}^{\dagger})$ which induces the energy
transition in a bound system irrelevant for free mCPs considered here. 
Furthermore, the second term $H_{I2}$ which can be expressed
roughly as $(\hat{a} \hat{a} + \hat{a} \hat{a}^{\dagger} + \hat{a}^{\dagger} \hat{a} + \hat{a}^{\dagger} \hat{a}^{\dagger})$ induces two photons transition. Only the second and the third term are relevant for our study since the
first and the last term violate photon number and energy
conservation for free particle system. Thus, we arrive at the following interaction term 
\begin{align}
\label{eq:HintF}
H_{I} \equiv \hat{H}_{\text{int}} &= \sum_{s} \frac{q^{2}_{s}}{2\,m_{s}} \, \left[ \frac{\hbar}{2 \,\epsilon_{0}\, \omega L^{3}} \right] 2\left(\hat{a}^{\dagger} \hat{a} + \frac{1}{2} \right)\,,\nonumber \\
 &= \frac{\epsilon^{2} \, e^{2}}{m_{\text{Q}}}\,\left[ \frac{\hbar\,\omega^{2}}{16\,\pi^{3} \,\epsilon_{0}\, c^{3}} \right] \left(\hat{a}^{\dagger} \hat{a} + \frac{1}{2} \right)\, N_{\text{Q}}\,, 
\end{align} 
where we have assumed all mCPs have the same charge $q_{s} = \epsilon\,e$ and the same mass $m_{s} = m_{\text{Q}}$ such that
the sum over mCPs is proportional to the total number of mCPs $N_{Q}$. In the second line of Eq.\eqref{eq:HintF}, we have substituted $L = 2\pi / k$ for $n = 1$ or single mode field. Such approximation is known as Jaynes-Cummings model in atomic transition~\cite{Jaynes:1963zz,Garrison:2008jh}. Experimentally, this is realized by
incorporating frequency selective component such as Fabry-Perot etalon into the optical cavity which picks a single mode with the lowest loss~\cite{Fox:2006quantum}.
\begin{figure}
	\centering
	\includegraphics[width=0.9\textwidth]{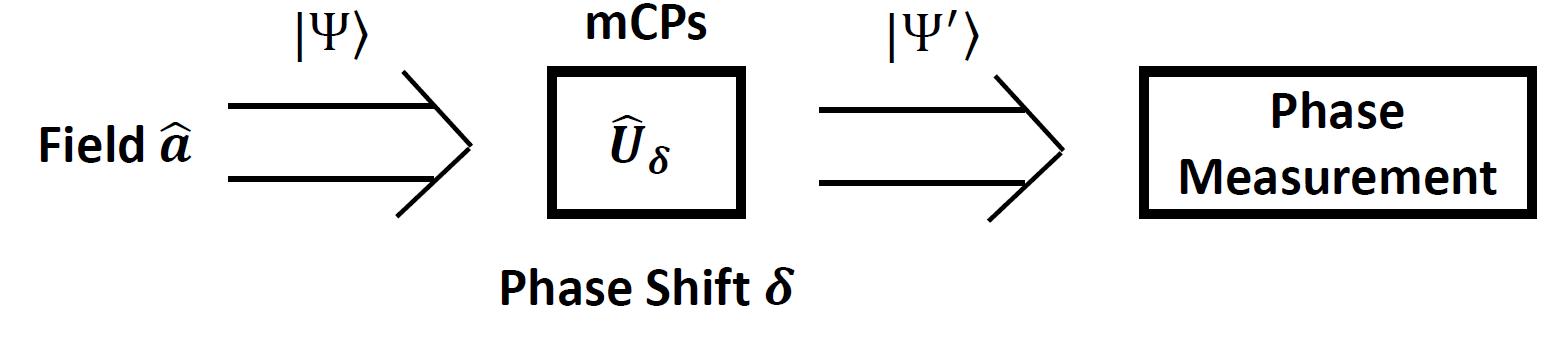}
	\caption{The phase shift $\delta$ of the optical field $\hat{a}$ changes the photon state from $\ket{\Psi}$ to $\ket{\Psi^{'}}$ due to photon-mCPs interaction.}
	\label{fig:interfero}
\end{figure}

We would
like to detect the phase shift
induced by the mCPs-photon interaction using
the phase measurement scheme illustrated in
Fig.~\ref{fig:interfero}. If we denote the initial photon state as $\ket{\Psi}$ and the final state post interaction as $\ket{\Psi^{'}}$, then they are related by unitary transformation $\hat{U}_{\delta}$~\cite{Kartner:1993,Ou:2017}
\begin{align}
\label{eq:psiP}
\ket{\Psi^{'}} &= \hat{U}_{\delta} \, \ket{\Psi} =e^{-\text{i}\,\hat{H}_{\text{int}}\text{t}/\hbar}\, \ket{\Psi}= e^{-\text{i}\,\hat{N}\delta}\,\ket{\Psi} 
\end{align}
where $ \hat{N} \equiv \hat{a}^{\dagger}\hat{a}$ is the photon number operator with the corresponding average value
$ N \equiv \left\langle \hat{a}^{\dagger}\hat{a} \right\rangle\gg 1$. 
Due to the quantum nature of the light, there is a
limitation that prevents us to probe the phase shift in arbitrary manner. According to quantum mechanics, the minimum detectable phase is given by the Heisenberg
limit~\cite{Dirac:1927, Heitler:1954}
\begin{align}
\label{eq:HeisLimit}
\Delta \delta \geq \frac{1}{N}\,,
\end{align}
where $N$ is the total number of photon. This limit is more sensitive than conventional laser interferometer which employ coherent state in its input. In this case, the detectable phase shift for $N$ number of photon is $\Delta \delta \geq 1/N^{-1/2}$ known as the standard quantum limit (SQL). There are several experiments that surpass the SQL in
the laboratories \cite{Bondurant:1984,Grangier:1987,Xiao:1987}. Typically,
the number of photon in modern laser interferometer is of the
order of $10^{20}$ or larger \cite{LIGOScientific:2016aoc} allowing
us to detect a minuscule phase shift in the lab.

\section{Phase Measurement Scheme}
\label{sec:phase}
Experimentally, a conventional photo-detection experiment on
the photon field $\hat{a}$ would
not allow us to extract its phase. To overcome this issue, one
needs to transform the photon state into phase sensitive
state. This is achievable via another unitary transformation $\hat{U}_{ps}$ on the initial photon field. If we denote the intial photon
state by $\ket{\Phi}$ and its final state after experiencing the phase
shift by $\ket{\Phi^{'}}$, one can construct the phase sensitive state
\begin{align}
\label{eq:Ups}
\ket{\Psi} = \hat{U}_{ps}\,  \ket{\Phi},\,\, \ket{\Psi^{'}}= \hat{U}_{ps} \,\ket{\Phi^{'}}\,,
\end{align}
where the detection of $\ket{\Psi^{'}}$ would give a
distinguishable
result from $\ket{\Psi}$. Since both of these states are phase
sensitive, one can probe the difference between them to
obtain the phase shift information. One of the best photon
states suitable for this task is the vacuum state. In practice, one may employ this state as  
$\ket{\Psi}$ such that the phase measurement on it would give zero result. A non-zero phase shift is implied when photons
are detected, which changes the corresponding vacuum state
into $\ket{\Psi^{'}}$. 

\begin{figure}
	\centering
	\includegraphics[width=0.9\textwidth]{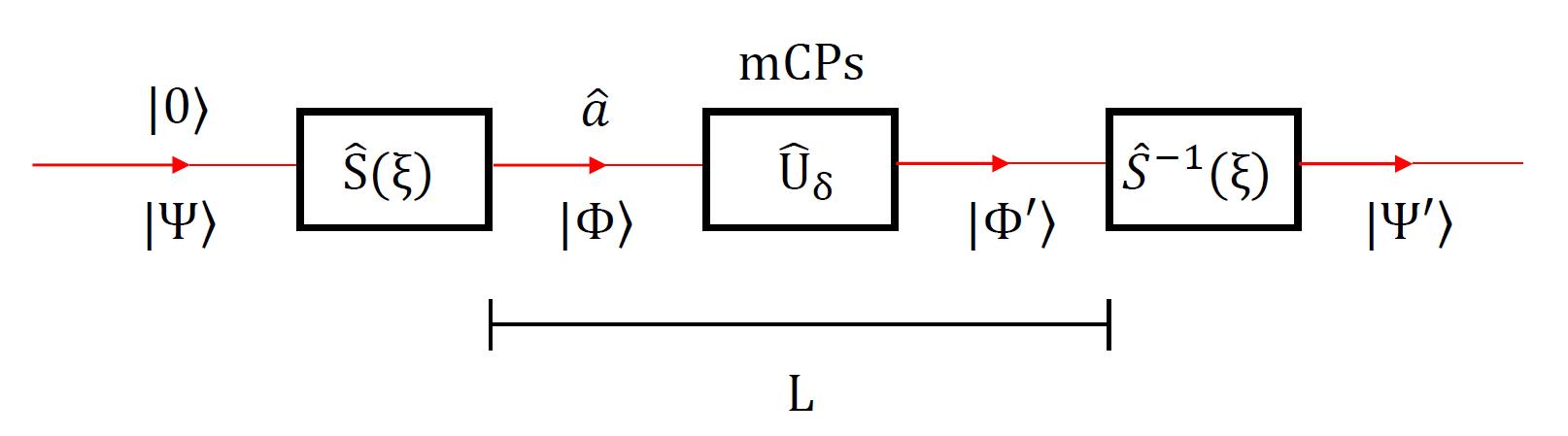}
	\caption{Single mode interferometer~\cite{Ou:2017,Yurke:1986,Nugroho:2023cun} with vacuum state as the input. The detected photons at the output implies non-vanishing phase shift $\delta$ from mCPs-photons interaction.}
	\label{fig:SMInterferometer}
\end{figure}

This mechanism can be implemented in a laser interferometer. However, in conventional interferometer comprising of two arms such as
Michelson interferometer, mCPs would interact with the
photons on both of arms leading to zero phase shift at the output. To
tackle this problem, we propose to employ unconventional
interferometer shown in Fig.\ref{fig:SMInterferometer}. This interferometer was proposed more than three decades ago \cite{Yurke:1986} and has been realized in the laboratories~\cite{Hudelist:2014,Manceau:2016esq,Liu:2018ahw,Xiao:2019,Ferreri:2020pxa}. Here, in
contrast to the traditional interferometer, the beam splitter is
replaced by the squeezing operator 
\begin{align}
\label{eq:defsqueeze}
\hat{S}(r) = e^{r (\hat{a}^{\dagger\,2} - \hat{a}^{2})/2}\,\,\, (r = \text{real}),
\end{align}
to form a single arm and single mode interferometer with the
squeezed vacuum state $\ket{\Phi} = \hat{S}(r)\, \ket{0}$
acting as the probe of the phase shift $\delta$~\cite{Ou:2017}.  
When acting on the field operator $\hat{a}$ and its corresponding Hermitian conjugate $\hat{a}^{\dagger}$, the squeezing operator would transform them into
\begin{align}
\label{eq:OpProperties}
\hat{S}^{\dagger}(r)\, \hat{a}\, \hat{S}(r) &= \hat{a}\, \text{cosh}\,r \, + \hat{a}^{\dagger} \, \text{sinh}\,r\,, \\
\hat{S}^{\dagger}(r)\, \hat{a}^{\dagger}\, \hat{S}(r) &= \hat{a}^{\dagger}\, \text{cosh}\,r \, + \hat{a} \, \text{sinh}\,r\,.
\end{align}
Therefore, in our phase measurement scheme, the corresponding phase sensitive operator as well as phase sensitive state are given by
\begin{align}
\label{eq:psState}
\hat{U}_{ps} = \hat{S}^{-1}(r) \,\,\, \text{and}\, \ket{\Psi} = \ket{0}\,.
\end{align}
Our proposal to detect earth-bound mCPs is shown in Fig.\ref{fig:SMInterferometer}. It is worth to mention that the
implementation of this setup needs to be realized in vertical
direction underground at the depth L (the interferometer arm length) from the surface.
In the absence of the phase shift, the output state would read $\ket{\Psi^{'}}_{\delta = 0} = \hat{S}^{-1}(r) \hat{S}(r) \ket{0} = \ket{0}$ or the vacuum state. On the other hand, in the presence of non-zero phase shift $\delta$, the output state becomes
\begin{align}
\label{eq:outState}
\ket{\Psi^{'}} = \hat{S}^{-1}(r)\, e^{-\text{i}\,\hat{a}^{\dagger} \hat{a}\, \delta}\, \hat{S}(r) \ket{0}\,.
\end{align}
As a result, the output port of the interferometer would read the average photon number given by
\begin{align}
\label{eq:nout}
\bra{\Psi^{'}} \hat{a}^{\dagger} \hat{a} \ket{\Psi^{'}} = 4\left\langle \hat{N} \right\rangle  \left[ 1 + \left\langle \hat{N} \right\rangle \right]\, \text{sin}^{2} \delta\,,
\end{align}
where $\left\langle \hat{N} \right\rangle = \text{sinh}^{2} r$ denotes the
average photon number evaluated with respect to the state $\ket{\Phi}$.

Taking
the noise at the output port as one photon, the signal-to-noise ratio (SNR) of our setup can be calculated as~\cite{Ou:2017}
\begin{align}
\label{eq:SNR}
\text{SNR} = 4\left\langle \hat{N} \right\rangle  \left[ 1 + \left\langle \hat{N} \right\rangle \right]\, \text{sin}^{2} \delta\,.
\end{align}
The noise in our SNR calculation originates from quantum
fluctuation of the light. Note that in the limit of $\left\langle \hat{N} \right\rangle >> 1$ and $\delta << 1$,
we arrive at the Heisenberg limit $\delta \sim 1/\left\langle \hat{N} \right\rangle$ for SNR equals to 1. We neglect other noise components such as photon loss and imperfect particle detection as
they can be mitigated using current
technology~\cite{Szigeti:2017}. In fact, the Heisenberg limit had been achieved in
laboratories~\cite{Linnemann:2016, Daryanoosh:2018, Anderson:2017} leaving only quantum noise relevant for SNR evaluation. It has been demonstrated that the "pumped-up" scenario in unconventional interferometer considered here would approach the Heisenberg limit with $10^{4}$ number of photons~\cite{Szigeti:2017}.
Furthermore, the sensitivity exceeding the Heisenberg limit ($\delta \geq 1/N^{-3/2}$), the so called "super-Heisenberg limit", had been achieved in the laboratory using $\mathcal{O} (10^{7})$ photons \cite{Napolitano:2011, Napolitano:2011b}. We expect such state-of-the-art could be implemented in
unconventional interferometer considered here in near future.  
From conservative point of view, we only consider the SQL and Heisenberg limit in our projected sensitivity below.

\section{Results and Discussion}
\label{sec:result}
\begin{figure}
	\centering
	\subfigure{{\includegraphics[width=6.6cm]{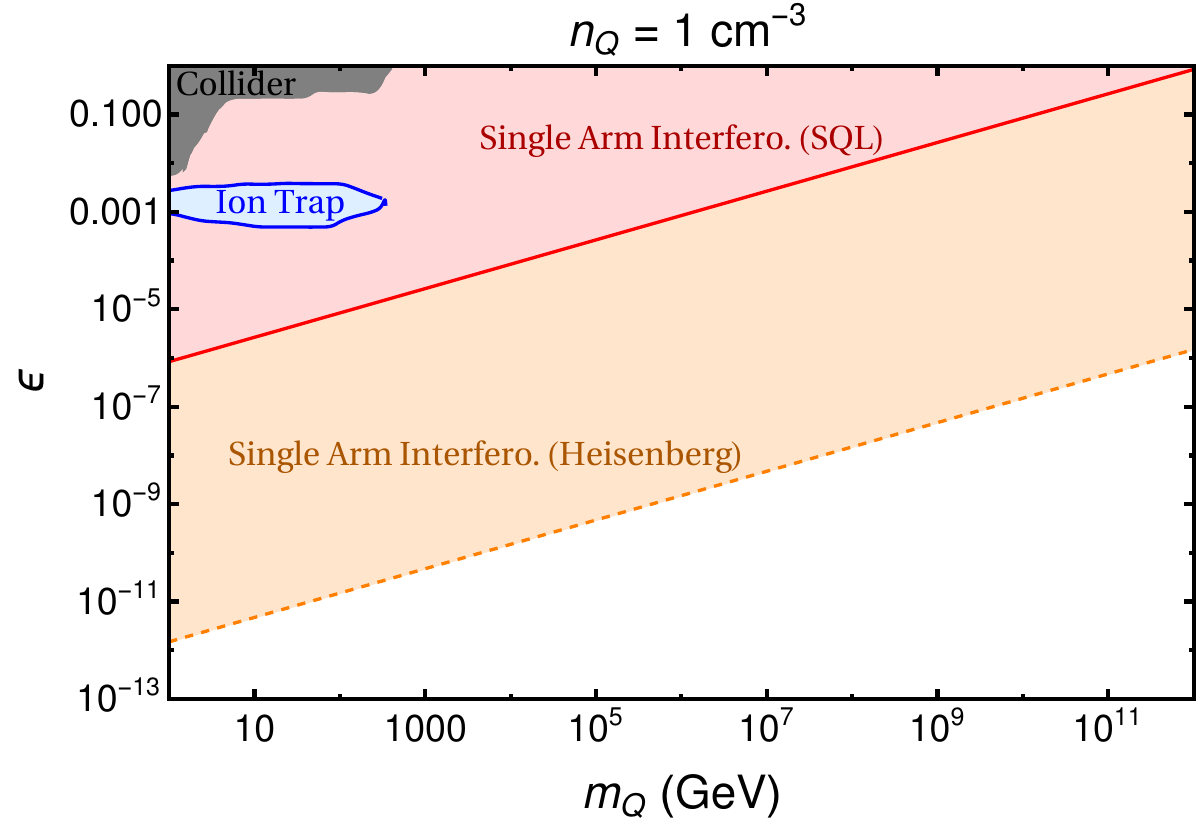} }}
    \subfigure{{\includegraphics[width=6.6cm]{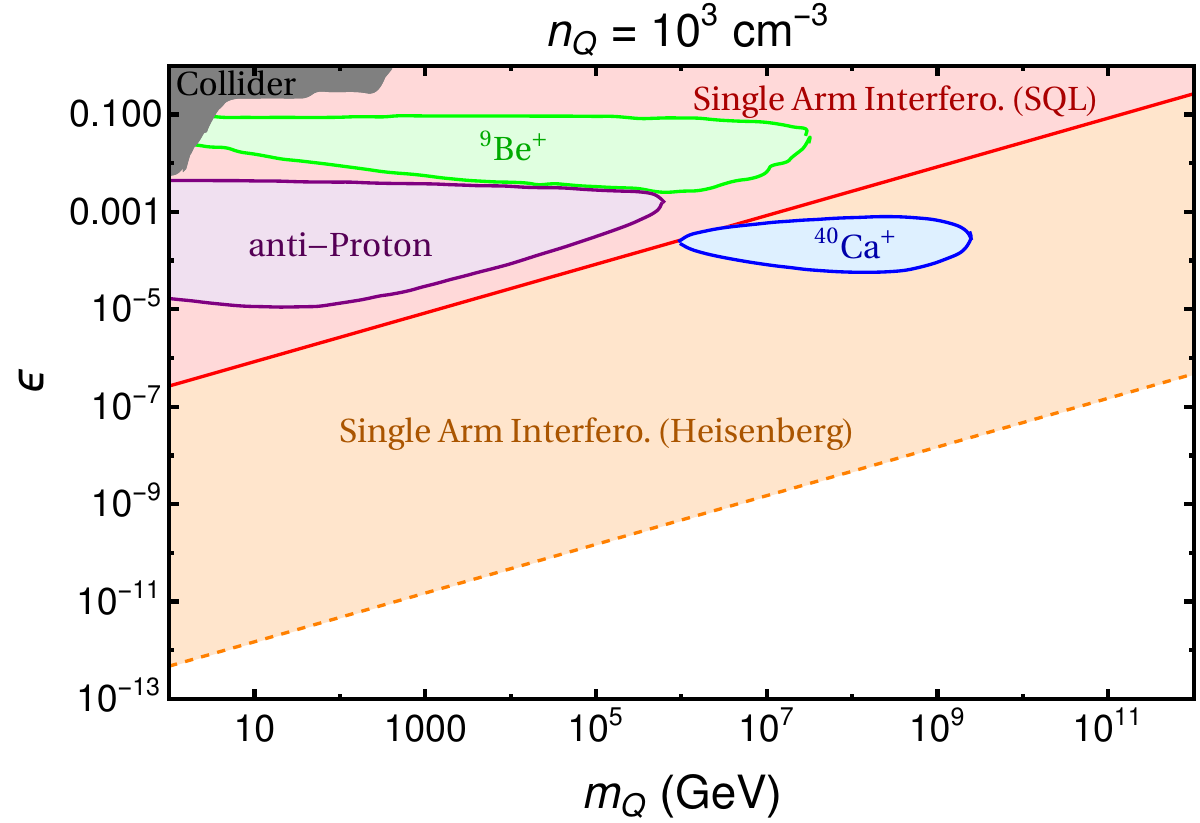} }}
    \subfigure{{\includegraphics[width=6.6cm]{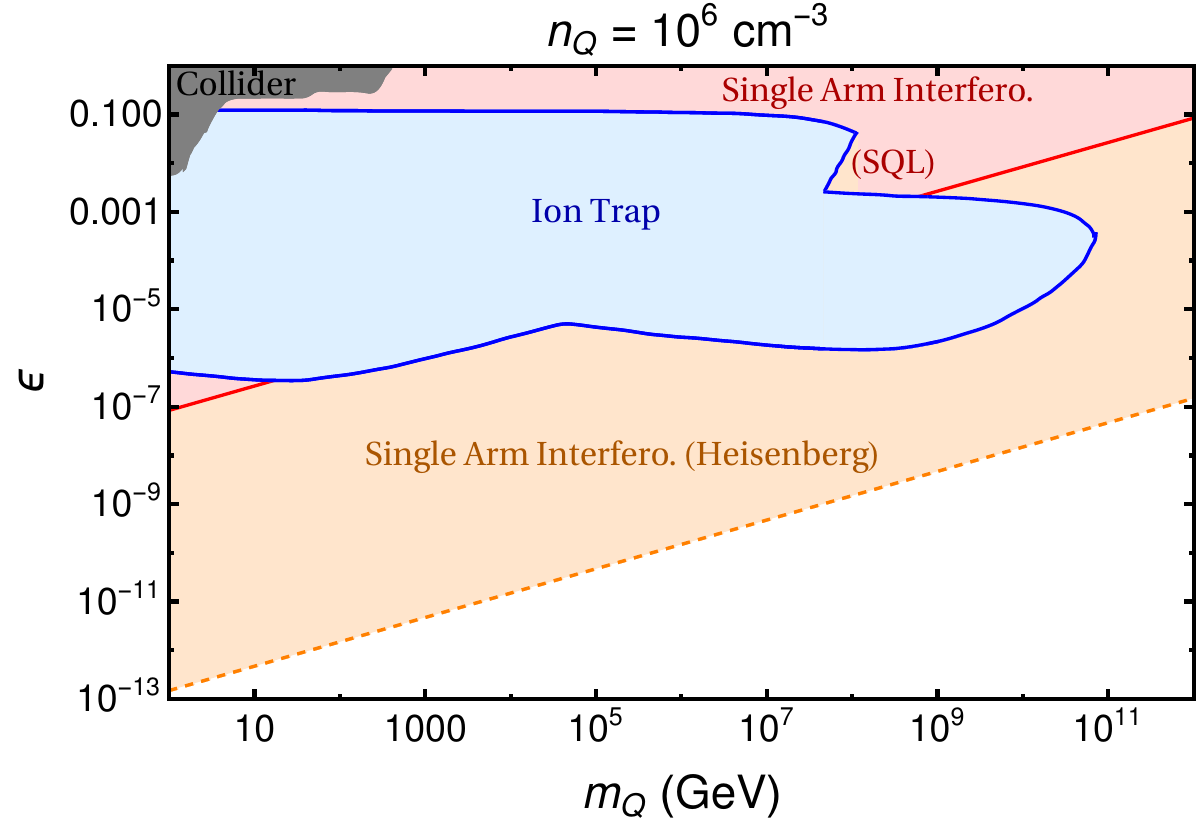} }}
	\caption{The projected sensitivities of single arm with single mode interferometer with arm length L = 1 km and 1064 nm laser for different mCPs densities $n_{\text{Q}}$: $1\,\text{cm}^{-3}$ (top-left), $10^{3}\,\text{cm}^{-3}$ (top-right), and $10^{6}\,\text{cm}^{-3}$ (bottom). We take the probe photon number $N = 10^{23}$~\cite{GammeVT-969:2007pci,Bahre:2013ywa,ALPS:2009des,Inada:2013tx} to detect mCPs induced phase shift.} 
	\label{fig:sensitivity}
\end{figure} 

To compute the SNR, one needs the explicit expression of the phase shift which is extracted from Eq~\eqref{eq:HintF} and \eqref{eq:psiP}
\begin{align}
\label{eq:delta}
\delta = \frac{\epsilon^{2} \, e^{2}}{m_{\text{Q}}}\,\left[ \frac{\omega^{2}}{16\,\pi^{3} \,\epsilon_{0}\, c^{3}} \right] \, N_{\text{Q}}\,t\,.
\end{align}  
Furthermore, we set the SNR value to be greater than
one to ensure that the noise would not exceed the hypothetical mCPs signal. Following~\cite{Budker:2021quh}, we consider three benchmark values of mCPs density $n_{\text{Q}}$: $1\,\text{cm}^{-3}$, $10^{3}\,\text{cm}^{-3}$, and $10^{6}\,\text{cm}^{-3}$ to get the projected sensitivities of our detection proposal.

The total mCPs number $N_{\text{Q}}$ that interacts with photons can be obtained by integrating the number of mCPs per unit length along the path traversed by the photons $\ell$ 
\begin{align}
\label{eq:ell}
N_{\text{Q}} = \int^{\text{L}}_{0} d\ell\, \tilde{n}_{\text{Q}}\,,
\end{align}
where L is the interferometer arm length and $\tilde{n}_{\text{Q}} = n^{1/3}_{\text{Q}}$ denotes the number of mCPs per unit length in $\text{cm}^{-1}$.
Here, the number of mCPs per unit
length for three different mCPs densities are $1\,\text{cm}^{-1}$, $10\,\text{cm}^{-1}$, and $100\,\text{cm}^{-1}$, respectively. Note that the
time parameter $t$ appearing in Eq.~\eqref{eq:delta} corresponds to the mCPs-photon
interaction
time $t = \text{L}/c$. Here, we set the arm length L equals
to 1 km, which is the typical length employed at GW experiment
such as LIGO~\cite{LIGOScientific:2016aoc}.

In Fig.~\ref{fig:sensitivity}, the projected sensitivities of single
arm interferometer are shown by the red line (SQL) and the
dashed orange line (Heisenberg limit) for different mCPs
densities. The corresponding light-red and light-orange shaded
area above the red line and dashed orange line show the parameter space probed by the interferometer 
operating at SQL and Heisenberg limit, respectively. The present
bounds from collider experiments are given by the gray area. For $n_{\text{Q}} = 1\,\text{cm}^{-3}$ in the upper-left panel, the
projected sensitivity of the ion trap proposal is displayed by the light
blue region. We see that our proposal is several order magnitudes more
sensitive than both collider and ion trap experiments, reaching the value of $\epsilon$ in the range of $10^{-11} - 10^{-6}$  for $1 \,\text{GeV} < m_{Q} < 10^{12}$ GeV assuming that the interferometer operates at Heisenberg limit. 
For $n_{\text{Q}} = 10^{3}\,\text{cm}^{-3}$, ion trap proposal covers different region of
parameter space 
depending on the ions employed in the trap as can be seen from the upper-right panel of Fig.~\ref{fig:sensitivity}. Still, our proposed setup is more sensitive reaching $\epsilon = 10^{-12}$ for 1 GeV mCP mass. As the number of mCPs
density increases, the ion trap proposal could probe the fractional charge $\epsilon$ as low as $10^{-6}$ in a wide range of mCPs mass (lower panel). 
In this case, our unconventional interferometer is taking a lead on the
sensitivity, covering the fractional charge $\epsilon$ between $10^{-13}$ to $10^{-7}$ for the same mass interval.
In all these figures, we see that even when the interferometer operates at SQL, the corresponding sensitivity reach is stronger than the present collider bounds. In addition, it also covers the parameter space given by the ion trap proposal. 
This shows that the single arm with single mode field   interferometer
can be utilized to detect earth-bound mCPs with $m_{Q} > 1$ GeV.
    
\section{Summary and Conclusion}
\label{sec:Summary}

Thanks to their non-vanishing EM interaction, mCPs with mass larger than 1 GeV are slowed down
and even stopped inside the earth. Consequently, their number
density undeground could reach several orders magnitude higher than
that of local virial DM. 
Since they interact with photons, we propose to detect their
existence by using unconventional interferometer. We employ single
arm interferometer with a squeezed vacuum state as the probe to
detect the induced phase shift from mCPs-photons interaction. To
gain a significant mCPs number density, this interferometer should be
placed deep underground. As an illustration, we set the
interferometer arm length to 1 km and place it in vertical direction below the earth's surface. 

We take three
benchmark values of mCPs density $n_{\text{Q}}$: $1\,\text{cm}^{-3}$, $10^{3}\,\text{cm}^{-3},\, \text{and} \, 10^{6}~\rm cm^{-3}$ to demonstrate the projected sensitivities of
our proposal. 
We calculate the SNR
for the mCPs induced phase shift for these different mCPs
densities and find out that the mixing parameter $\epsilon$
can be probed as low as $10^{-11}\,,10^{-12}$, and $10^{-13}$,
respectively, for mCPs mass around 1 GeV assuming that the interferometer works at Heisenberg limit. As
compared with the current bound from the LHC, where $\epsilon$ is of
order $10^{-3}\sim10^{-2}$, the sensitivity of our proposal is
about $10^{10}$ higher. Moreover, the projected sensitivities
of our setup surpass the recently proposed mCPs detection
using ion traps~\cite{Budker:2021quh} in a wide
mCPs mass range between 1 GeV to $10^{12}$ GeV. All in all, we have demonstrated that single
arm interferometer with single mode photon field provides a
suitable venue to detect earth-bound mCPs with unprecedented reach of sensitivity.

\section*{Acknowledgment}
\label{sec:Acknowledgment}
This work was supported by the National Science and Technology Council (NSTC) of Taiwan under Grant No. NSTC 112-2811-M-003-004- and NSTC 113-2811-M-003-019.

\end{document}